\documentclass[twocolumn,prl,showpacs,preprintnumbers,amssymb]{revtex4}

\usepackage{epsfig}
\usepackage{dcolumn}
\usepackage{bm}

\def\IC{\bf C}
\def\IZ{\bf Z}

\def\z2z2{$\IC^3/(\IZ_2\times\IZ_2)$}

\def\k{\kappa}

\def\th{\theta}

\def\beq{\begin{equation}}\def\eeq{\end{equation}}
\def\beqa{\begin{eqnarray}}\def\eeqa{\end{eqnarray}}
\def\barr{\begin{array}}\def\earr{\end{array}}

 \let\br=\bigr

\def\bd{\begin{document}}
\def\ed{\end{document}}
\def\ba{\begin{array}}
\def\ea{\end{array}}
\def\bea{\begin{eqnarray}}
\def\eea{\end{eqnarray}}
\def\ft#1#2{{\textstyle{{\scriptstyle #1}\over {\scriptstyle #2}}}}
\def\fft#1#2{{#1 \over #2}}
\newcommand{\be}{\begin{equation}}
\newcommand{\ee}{\end{equation}}
\newcommand{\eq}[1]{(\ref{#1})}
\def\eqs#1#2{(\ref{#1}-\ref{#2})}
\def\det{{\rm det\,}}
\def\tr{{\rm tr}}
\newcommand{\ho}[1]{$\, ^{#1}$}
\newcommand{\hoch}[1]{$\, ^{#1}$}
\def\ra{\rightarrow}

\def\Xh{\hat{X}}
\def\ah{\hat{a}}
\def\xh{\hat{x}}
\def\yh{\hat{y}}
\def\ph{\hat{p}}
\def\G{{\cal G}}
\def\Dth{{\Delta_\th}}

\def\bk{{\bf k}}
\def\bx{{\bf x}}
\def\br{{\bf r}}
\def\tr{{\rm tr \,}}
\def\Tr{{\rm Tr \,}}
\def\diag{{\rm diag \,}}
\def\tg{{\rm tg \,}}
\def\NPB#1#2#3{Nucl. Phys. B {\bf #1} (19#2) #3}
\def\PLB#1#2#3{Phys. Lett. B {\bf #1} (19#2) #3}
\def\PLBold#1#2#3{Phys. Lett. {#1B} (19#2) #3}
\def\PRD#1#2#3{Phys. Rev. D {\bf #1} (19#2) #3}
\def\PRL#1#2#3{Phys. Rev. Lett. {\bf #1} (19#2) #3}
\def\PRT#1#2#3{Phys. Rep. {\bf #1} C (19#2) #3}
\def\MODA#1#2#3{Mod. Phys. Lett.  {\bf #1} (19#2) #3}
\def\ov{\overline}

\begin{document}

\preprint{CU-TP-1039, UPR-952-T}

\title{Imprints of Short Distance Physics On Inflationary Cosmology }

\author{Richard Easther$^{1}$, Brian R. Greene$^{1,2}$, William H.
Kinney$^1$ and Gary Shiu$^{3}$}
\affiliation{$^1$Institute for Strings, Cosmology and Astroparticle
Physics, Columbia University, New York, NY 10027, USA \\
$^2$Department of Mathematics, Columbia University, New York, NY
10027, USA \\ $^3$Department of Physics and Astronomy, University of
Pennsylvania, Philadelphia PA 19104-6396, USA}

\begin{abstract}
We analyze the impact of certain modifications to short distance
physics on the inflationary perturbation spectrum.  For the specific
case of power-law inflation, we find distinctive -- and possibly
observable -- effects on the spectrum of density perturbations.
\end{abstract}
\maketitle


Inflation stretches quantum fluctuations to astrophysical scales,
providing a microscopic mechanism for the formation of galaxies
\cite{structure}. Most models of inflation yield far more
quasi-exponential expansion than the 60 e-foldings required to solve
the difficulties faced by the standard model of the hot big bang.
Consequently, astrophysical scales in the present universe map to
physical distances in the primordial universe that are exponentially
smaller than any conceivable fundamental length. As explained in
\cite{brand1,BM,N}, this introduces an implicit assumption into the
perturbation spectrum calculation: that spacetime physics and quantum
mechanics can be extrapolated to arbitrarily small physical lengths,
independently of any fundamental length scale.  A fundamental length
is predicted by almost all attempted unifications of general
relativity with the other fundamental forces of nature, and also by
quantum theories of gravity. Na{\"\i}ve dimensional arguments identify
this scale with the Planck length but, for example, the additional
scale introduced by string theory, the string length, can easily be
one or two orders of magnitude larger than the Planck length
\cite{stringscale}, if not more \cite{adds}.  Thus we ask two
important questions: can a fundamental length change the predicted
perturbation spectrum and, if so, are these differences detectable
observationally?  If these questions are answered in the affirmative,
they may provide astrophysical tests for theories of nature containing
a fundamental length scale, including string theory.

Initial investigations \cite{BM,N} relied on assessing the robustness
of the usual inflationary spectra to changes in the perturbations'
dispersion relations.  A different approach was initiated in
\cite{Kempf,KN,EGKS} where a nominally string inspired fundamental
length appears in the uncertainty relations and provides a short
distance cutoff.  We showed that the de Sitter space power spectrum
predicted for this model is rescaled by a multiplicative constant,
observable only if one has independent knowledge of the Hubble
constant during inflation \cite{EGKS}.  We emphasize that the
calculations of \cite{EGKS} and those presented here are not performed
within string theory, but instead make use of a standard field theory
modified on short scales in a manner inspired by string theory.  Our
aim is to determine if such short distance scale modifications can
yield astrophysically observable signatures.

In almost all non de Sitter inflationary backgrounds the expansion
rate is slower than exponential, and the physical horizon size
increases with time. We predicted that the spectrum will be changed more
dramatically at long wavelengths than at short wavelengths (larger
$k$), since the impact of the fundamental length increases with its
ratio to the physical horizon size \cite{EGKS}. Since the resulting
change to the spectrum is more complicated than a simple rescaling,
it is -- in principle -- observable.

This letter discusses the perturbation spectrum generated by power law
inflation when a fundamental length is inserted into the uncertainty
relations.  The shape of the spectrum does change, in line with our
initial expectation. To this extent, we agree with \cite{KN} which
presents a qualitative analysis of the same problem.  However, the
magnitude of the effect drops more slowly as the fundamental length
decreases than the analysis of \cite{KN} suggests, with important
consequences for the possibility of observing this signal.

In broad outline, the procedure for determining the spectrum is not
changed by the introduction of a fundamental length. We begin with
appropriately normalized field oscillations, which are quantum in
origin but obey the classical equations of motion, and extract the
spectrum from their asymptotic amplitudes.  However, the fundamental
length modifies both the evolution equation and the explicit form of
the normalization condition.

For wavelengths much greater than the fundamental length, tensor modes
$v_k$ obey
\begin{equation}
v_k'' + \left(k^2 - {\frac{a''}{a}}\right) v_k = 0,
\end{equation}
with $P_g^{1/2} \propto \left|v_k / a\right|$. Scalar modes $u_k$ obey
a similar equation, with $a$ replaced by $z \equiv a \dot\phi /H$.
Here $\phi$ is the field responsible for inflation, and $P_s^{1/2}
\propto \left|u_k / z\right|$. For the special case of power-law
inflation, $z \propto a$, so scalar and tensor modes obey identical
equations, although their power spectra differ in their
normalizations.  Since the two types of modes have the same equations
of motion at long wavelength, we assume that they also obey the same
equations of motion at {\em short} wavelength, where the influence of
short-distance physics is important, and that any modulations of the
power spectrum due to short-distance physics apply identically to
tensor and scalar modes.  With this assumption, modifications to short
distance physics can result in violations of the so-called
``consistency condition'' for inflation\cite{hui01,Shiu:2001sy}. While
this is reasonable, it is by no means guaranteed: scalar modes are a
mix of field and metric fluctuations in an arbitrary gauge \cite{MFB},
where as the tensor modes are purely metric. It is not clear
that a short-distance cutoff will affect fluctuations of the metric in
the same way as fluctuations in an arbitrary scalar field. However,
general coordinate invariance implies that we can transform (for
example) to a gauge in which even the scalar fluctuations are purely
``metric'', and if this property is preserved at short distances then
the effect of new physics on the scalars and tensors should be
identical.  Even in the existing literature, metric (gravitational
wave) fluctuations can be treated as a generic scalar field at short
distances. This is also reasonable, but not inevitable.

Following \cite{Kempf}, tensor fluctuations
$v_{\tilde{k}}$ obey:
\begin{equation}\label{ukmode}
v_{\tilde{k}}^{\prime\prime} + \frac{\nu^\prime}{\nu}
v_{\tilde{k}}^\prime +\left(\mu 
 -  \frac{a^{\prime \prime}}{a} 
-\frac{a^\prime}{a} \frac{ \nu^\prime}{\nu}
\right) v_{\tilde{k}} = 0
\end{equation}
where $a$ is the scale factor, the prime denotes differentiation with
respect to conformal time $\eta$, while $\tilde{k}^i = a
\rho^i e^{-\beta \rho^2/2}$ with $\rho^i$ being the Fourier transform of
the physical coordinates $x^i$, and
\be
\mu(\eta,\rho)  \equiv  
\frac{a^2 \rho^2}{(1-\beta \rho^2)^2},\qquad
\nu(\eta,\rho)  \equiv  
\frac{e^{\frac{3}{2}\beta \rho^2}}{
 \left( 1 - \beta \rho^2 \right)}.
\ee
When evaluating the derivatives of $v_{\tilde{k}}$ with respect to
$\eta$, we are holding $\tilde{k}$ (and not the usual comoving
momentum $k$) fixed with time.  It is therefore convenient to express
$\mu$ and $\nu$ in terms of $\tilde{k}$ by introducing the Lambert $W$
function \cite{Lambert}, which is defined so that $W(xe^x)=x$:
\begin{equation}
\mu = -\frac{a^2}{\beta} \frac{W(\zeta)}{\left( 1+W(\zeta) \right)^2}~,
\quad
\frac{\nu^{\prime}}{\nu} = 
\frac{a^{\prime}}{a} \frac{W(\zeta) \left(5+3 W(z \right))}{(1+W(\zeta))^2}~.
\end{equation}
where $\zeta=-\beta \tilde{k}^2/a^2$.

For power law inflation
\begin{equation}
a(t) = t^p, \qquad a(\eta) = \left(\frac{\eta}{\eta_0}\right)^q,
 \qquad q= \frac{p}{1-p}.
\end{equation}
For inflation to occur, we need $p>1$.

The cutoff is introduced by requiring that $\rho^2 \leq 1 / \beta$,
motivated by the notions of a minimum distance in string theory and
the so-called ``stringy uncertainty principle" \cite{sur}.  Fluctuations
with comoving wavenumber $k$ reach the cutoff $\rho^2 = 1/\beta$ at
$\eta_k$, where
\be
\eta_k =  \eta_0 \left(  e \beta \tilde{\kappa}^2 \right)^{1/2q} 
            = \frac{1}{1-p} \left( e \beta 
\tilde{\kappa}^2\right)^{1/2q}
\ee
where the implicit definition of $\eta_0$ comes from setting $a(t)=1$
when $t=1$, and $t$ is the usual physical time.

Writing $\eta =\eta_k (1-y)$ in order to extract the $k$ dependence
and abbreviating $W(\zeta)$ as $W$, we have
\begin{eqnarray}
 && \ddot{v}_{\tilde{k}}  -
\frac{q}{1-y} \frac{W \left(5+3 W\right)}{(1+W)^2} 
\dot{v}_{\tilde{k}} - \left(
\frac{(1-y)^{2q} \eta_k^{2q+2}  W}{\beta \eta_0^{2q}\left(
1+W\right)^2}  \right. \nonumber \\  
&& \left. + \frac{q(q-1)}{(1-y)^2} + \frac{q^2}{(1-y)^2}
\frac{W \left(5+3 W \right)}{(1+W)^2} 
\right) v_{\tilde{k}} = 0
\end{eqnarray}
In the de Sitter case, $k$ can be eliminated from the equation of
motion, but this cannot be done here. During power-law inflation,
different modes sample a different value of the Hubble constant as
they cross the horizon, and this will be reflected in the
scale-dependent modifications to the spectrum we will observe. Despite
this, our analysis of the de Sitter case \cite{EGKS} can easily be
adapted to the power-law problem.

When $\zeta= -1/e$, $W(\zeta)$ has a branch point \cite{Lambert}. Physically,
this is the moment when $y=0$ ($\eta=\eta_k$) and the fluctuation
with wavelength $k$ is ``created''.  As in \cite{EGKS}, we solve for
the leading behavior of $v_{\tilde{k}}$ by extracting the most
singular terms of the equation of motion,
\begin{equation}\label{ukmodey}
\ddot{v}_{\tilde{k}}
- \frac{1}{2y} \dot{v}_{\tilde{k}}
+\frac{A_k}{y} v_{\tilde{k}} = 0,
\end{equation}
where dots denote derivatives with respect to $y$, and
\begin{equation}
A_k = - \frac{1}{q} \frac{\eta_k^{2q+2}}{4 \beta \eta_0^{2q}} -
\frac{q}{2}.
\end{equation}

The solution to (\ref{ukmodey}) is:
\begin{equation}\label{uk0}
v_{\tilde{k}}^{(0)} (y) = y^{3/4} \left(C H_{-3/2}^{(2)}(2\sqrt{A_k
y}) + D H_{-3/2}^{(1)}(2\sqrt{A_k y}) \right).
\end{equation}
Here, $H_{-3/2}^{(2)}$ is the second Hankel function, $C$ and
$D$ are constants.  The first Hankel function is its complex
conjugate $H_{q/2-1}^{(1)}=H_{q/2-1}^{(2)*}$.

The solution is normalized by the Wronskian condition
\begin{equation}\label{wronskianu}
v_{\tilde{k}}(\eta) v_{\tilde{k}}^{* \prime} (\eta)- v_{\tilde{k}}^{*}
(\eta) v_{\tilde{k}}^{\prime} (\eta) = i \left(1-\beta \rho^2 \right)
e^{-\frac{3}{2} \beta \rho^2}.
\end{equation}
Using Hankel function identities  \cite{abramowitz} we deduce
\begin{equation}
|C|^2 -|D|^2 = -\eta_k \pi \sqrt{-q} e^{-3/2}. \label{constraint}
\end{equation}
This, together with equations~(\ref{uk0}) and (\ref{spectrumdef}),
gives the general result for all possible boundary conditions.  More
specifically, though, we must fix $C$ and $D$, {\it i.e.\/} we must
specify boundary conditions for equation~(\ref{ukmodey}) or,
equivalently, the form of the vacuum state.  Ultimately, a more
complete understanding of short scale physics would allow a first
principles selection of boundary conditions.  Here we simply note that
if $D\ne0$, the spectrum never approaches the exact power-law form,
even when the Hubble parameter is arbitrarily small.  Consequently,
$D=0$ is the only vacuum choice {\em with constant coefficients\/}
that reduces to the Bunch-Davies vacuum at low energies.  We thus
focus our analysis on the ``minimal'' choice that $D=0$ for all $k$,
but we stress that this choice is not unique: $C$ and $D$ could depend
on $\beta$, $k$ or $H$ in a such a way that the constraint of
(\ref{constraint}) was always satisfied, and that $D$ was non-zero at
high energies, and vanishing sufficiently rapidly as $\beta
\rightarrow 0$ to avoid any experimental constaints on non-zero $D$.
Our results obviously depend on this choice and this issue certainly
deserves further investigation.  Finally, we point out that that
setting $q=-1$ in Equation~(\ref{constraint}) reproduces the de Sitter
result, after reconciling the normalization factors.

We solve the mode equations numerically \cite{EGKS,Easther} and match
the numerical solution to the approximate analytical form, including
sub-leading corrections, near $y=0$.  We obtain the scalar spectrum by
solving the mode equations for multiple values of $k$, and then
extracting the necessary late time limit to compute
\be \label{spectrumdef}
P_g^{1/2} = \sqrt{\frac{k^3}{2 \pi^2}} \left| \frac{v_k}{a}
\right|_{k=aH}, \quad P_s^{1/2} = \sqrt{32 \pi p } P_g^{1/2},
\ee
where we have obtained the scalar spectrum from the tensor one, as
outlined above.

\begin{figure}[tbp]
\begin{center}
\begin{tabular}{c}
\epsfxsize=8cm 
\epsfbox{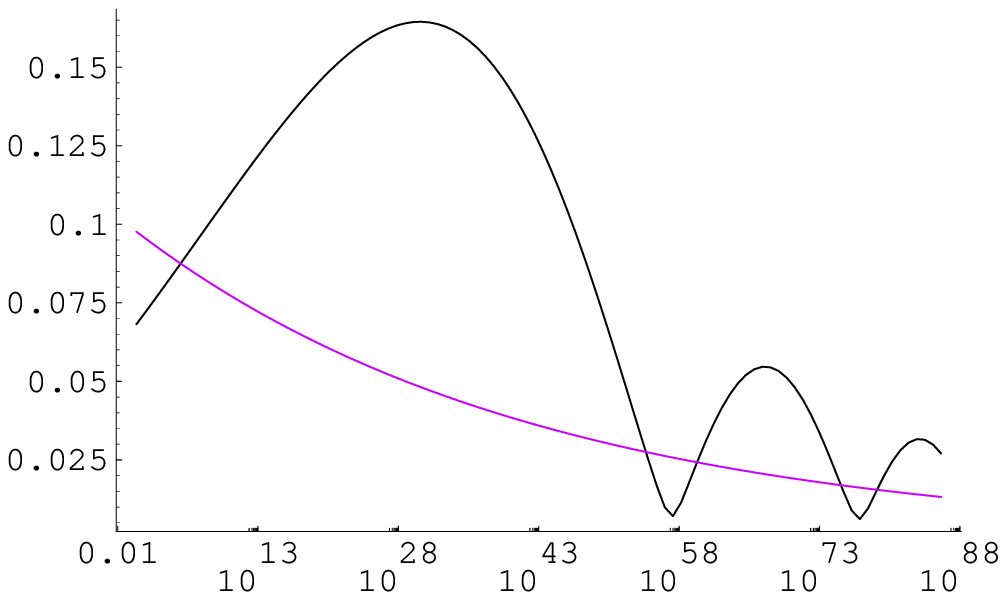} 
\end{tabular}
\end{center}
\caption[]{ The scalar spectrum, $P^{1/2}_s(k)$ is plotted against
$k$, with $\sqrt{\beta}=100$ and $p=100$, where $k=1$ corresponds to
$k_{\mbox{crit}}$. The standard power law spectrum is plotted for
comparison (smooth line).}
\end{figure}

\begin{figure}[tbp]
\begin{center}
\begin{tabular}{c}
\epsfxsize=8cm 
\epsfbox{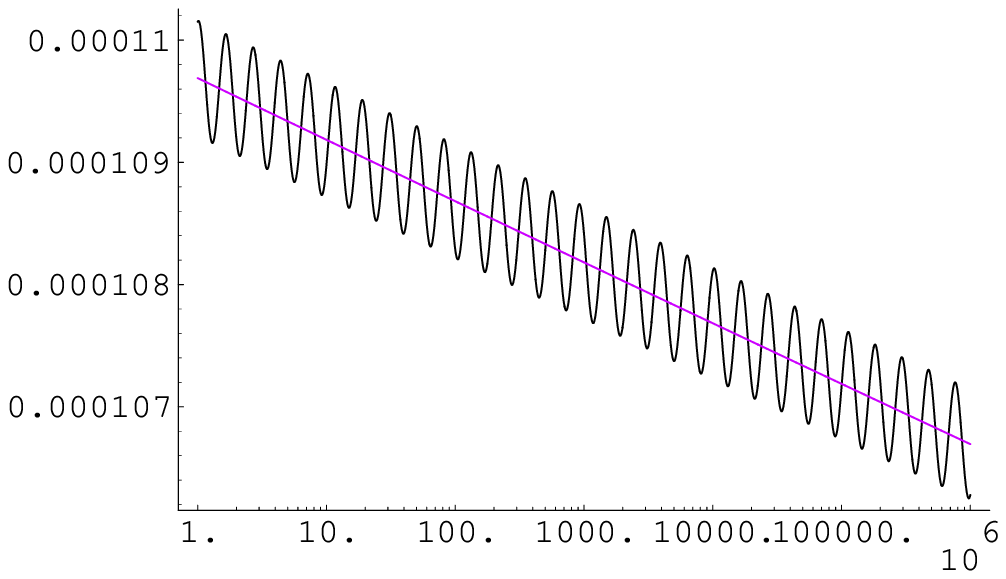} \\
\epsfxsize=8cm 
\epsfbox{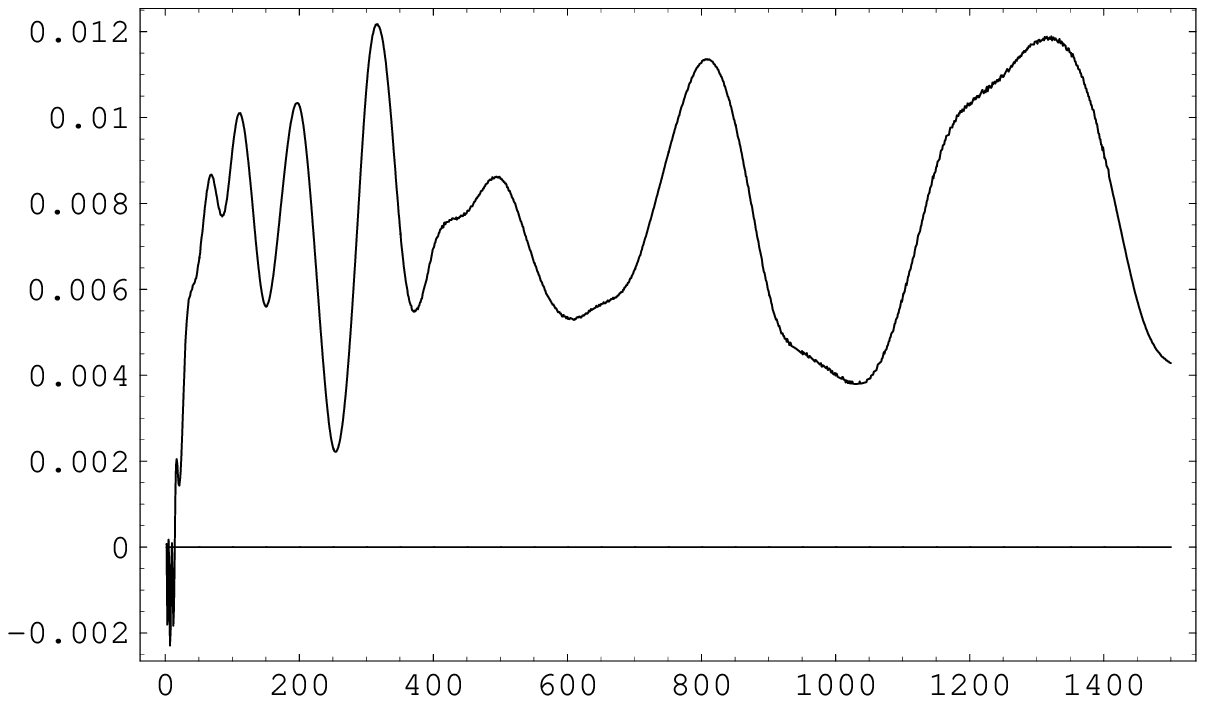}
\end{tabular}
\end{center}
\caption[]{ The top plot plots $P^{1/2}_s(k)$ against
$k$ for $\sqrt{\beta}=100$ and $p=500$, and with the straight line
showing the standard power-law result. The modulation of the spectrum
has an amplitude of $0.4$\% of the overall signal.  The bottom plot
shows the percentage change in the $C_l$ values (plotted against $l$)
computed from this spectrum, relative to the spectrum calculated in
the absence of a fundamental length.}
\end{figure}

The spectrum is only well defined if the minimum length ($\sqrt\beta$)
is less than the horizon size ($1/H$), or $\sqrt{\beta}H$ is less than
unity. The critical mode, $\k_{\mbox{crit}}$, that crosses the horizon
at the moment when $\sqrt{\beta}H = 1$ is
\begin{equation}
k_{\mbox{crit}} = p \left( e \beta p^2 \right)^{(p-1)/2}.	
\end{equation}  
For large values of $p$, $\k_{\mbox{crit}}$ is enormous. This reflects
the massive amount of inflation that takes place between the Planck
time ($t=1$ in natural units) and the moment at which $\sqrt{\beta}H =
1$; the numerical value of $k$ can always be rescaled by redefining
$a_0$, the value of $a$ when $t=1$.

Fig.~1 shows the spectrum for the longest modes, with $p=100$. There
is a large modulation in the spectrum, corresponding to the slow
decrease in $H$ as the universe evolves. However, these modes have a
much larger amplitude than those contributing to the CMB power
spectrum and structure formation. Fig.~2 displays the spectrum with
$p=500$ and a ``window'' of $k$ values with amplitudes of the same
order as the modes which are the precursors to structure formation. We
have not carefully normalized this spectrum (which requires
assumptions about the dark matter composition, $\Omega_\Lambda$, {\it
etc.\/}), since any signal of trans-Planckian physics is much
smaller than the uncertainty in currently available data. Instead we
assume that $10^{-6} \lesssim \Delta T/T \lesssim 10^{-5}$, which --
given that $(\Delta T/T)^2 \simeq P_s/180$ \cite{turner} -- corresponds to $1.5
\times 10^{-5} \lesssim P_s^{1/2} \lesssim 1.4 \times 10^{-4}$.

The standard power-law spectrum is modulated by an ``oscillation''
whose amplitude and wavelength depend on both the fundamental length
$\sqrt{\beta}$ and the power-law parameter, $p$. The oscillations are
attributable to successive modes undergoing increasing numbers of
periods between the initial time and horizon exit, with a full extra
period corresponding to a single oscillation in the spectrum.

The amplitude and period (in $\log{k}$) of the oscillations are
roughly proportional to $\sqrt{\beta}$.  In principle $\sqrt{\beta}$
is predicted by fundamental theory, but from our perspective here it
is a free parameter.  If $\sqrt{\beta}$ is identified with the string
scale, it could conceivably be two orders of magnitude longer than the
Planck length, and we use this value in the numerical plots.
Observationally, the key parameter is the ratio of the fundamental
length to the Hubble radius $\sqrt{\beta}H$. In power-law inflation
(and any other non de Sitter model) $H$ is a slowly changing
parameter. The observationally relevant range of $H$ is fixed by the
amplitude of the power spectrum, which is deduced from observations of
the CMB and large scale structure.

The rate of change of $H$ is determined by $p$, and as $p$ increases
the wavelength of the fluctuations in the spectrum increases while
their amplitude goes down.  This accords with our physical
understanding of the oscillations: for a fixed value of $H$ (and thus
$P^{1/2}_s$), $\dot{H}$ decreases with increasing $p$. Thus the
wavelength of the oscillations (in $\log{k}$) increases with $p$,
since $H$ at horizon exit changes more slowly with $k$ at larger $p$ .
The variation in the amplitude arises because we are effectively
holding $H$ fixed at horizon exit, but $H(\tau_k)$ decreases as $p$ is
increased. Consequently, the effective value of $\beta H^2$ for a
given mode decreases as $p$ is increased, which accounts for the $p$
dependence of the amplitude of the oscillations.  The oscillations do
not vanish as $p$ becomes arbitrarily large, although their wavelength
becomes arbitrarily long, and we approach the de Sitter limit where
the spectrum is shifted by a constant multiplicative factor.

In Fig.~2, $p=500$ the spectrum is almost flat, and the $C_l$ values
that would be measured by CMB experiments are modified by between $.5$
and $1$\%.  A signal of this size lies at the limits of detectability,
even with ideal experiments, and would be swamped by cosmic variance
at all but the largest values of $l$.  Existing constraints on the
spectral index put a weak lower bound on $p$ of around 20. With this
value, the oscillations' wavelength is so short that the resulting
spectrum appears to include a random noise term when plotted over the
range of $k$ values relevant to structure formation.

Despite the extreme challenge and perhaps near impossibility of
detecting an effect of this size for reasonable values of $\beta$, the
conclusions of this letter are still much more optimistic than we
might have otherwise expected.  First, and in accord with our previous
de Sitter calculation, we find that the magnitude of the modification
to the spectrum is a function of $(\beta H^2)^n$, where $n$ appears to
be slightly smaller than $1/2$. This disagrees with ~\cite{KN}, in
which it is argued that $n$ is roughly unity.  However, \cite{KN}
relies on a WKB approximation to the mode equation and chooses the
vacuum to be the purely ``$-$'' WKB solution.  We have decomposed our
numerical solutions into the two WKB solutions at a time when the WKB
approximation holds well, and the actual solution (using the initial
conditions described above and also advocated by \cite{KN}) contains a
mixture of both WKB solutions, where the coefficient on the plus
``$+$'' scales like $(\beta H^2)^n$ with $n\lesssim 0.5$.  It is this
mode mixing of the standard WKB solutions, with a mixing coefficient
of order $\sqrt{\beta}H$, that accounts for the magnitude of our
results. Thus the pure ``$-$'' WKB approximation is not consistent
with these initial conditions, perhaps explaining why the estimate of
\cite{KN} for the impact of the fundamental length on the spectrum is
significantly less than we find here.

We have assumed that the minimum length lies somewhat below the Planck
scale. While this is justifiable from a stringy perspective, if we had
put the fundamental length equal to the Planck length ($\sqrt{\beta} =
1$) the effect we see decreases significantly. Moreover, as mentioned,
the equations we solve are not derived from string theory and hence
there is no guarantee that they are the ones that quantum gravity will
give us.  {\em Nevertheless, we find it encouraging that whereas there are
16 orders of magnitude separating the Planck scale from conventional
accelarator experiments, the string scale modifications we study here
yield cosmological effects that may be only one or two orders of
magnitude below the threshold of observability.}

In principle existing CMB measurements put experimental restrictions
on a portion of the $(\beta,p)$ plane. Given the accuracy of current
data, the constraints on $\beta$ would be extremely weak, and we have
not performed this calculation. As CMB data and surveys of Large Scale
Structure (and our ability to work backwards from the observed to the
primordial spectrum) improve, it may become possible to place
meaningful restrictions on short scale physics using astrophysical and
cosmological data.

It is natural to ask whether we can do a full string theoretic version
of this calculation.  One approach would be to study the full
two-point functions of graviton and inflaton string excitations,
either from string field theory or along the lines of
Ref.~\cite{CMNP}.  These quantities are sensitive to the large number
of high energy degrees of freedom found in string theory, another
inherently stringy feature.

\acknowledgments

We thank Robert Brandenberger and Jens Niemeyer for discussions. The
work of BG is supported in part by DOE grant DE-FG02-92ER40699B and
the work of GS was supported in part by the DOE grants
DE-FG02-95ER40893, DE-EY-76-02-3071 and the University of Pennsylvania
School of Arts and Sciences Dean's funds. ISCAP gratefully
acknowledges the generous support of the Ohrstrom Foundation.


\begin{thebibliography}{99}

\bibitem{structure} 
V.F. Mukhanov, G.V. Chibisov, JETP Lett. {\bf 33} (1981) 532;
Sov. Phys. JETP {\bf 56} (1982) 258;  
S.W. Hawking, Phys. Lett. {\bf B115} (1982) 295; 
A.A. Starobinsky, {\it ibid}, {\bf B117} (1982) 175;
A.H. Guth, S.-Y. Pi, Phys. Rev. Lett. {\bf 49} (1982) 1110; 
J. Bardeen, P.J. Steinhardt, M. Turner, Phys. Rev. {\bf D28} (1983)
679; 
V.F. Mukhanov, JETP Lett. {\bf 41} (1985) 493.


\bibitem{brand1} R. Brandenberger,  hep-ph/9910410.


\bibitem{BM} 
J.~Martin and R.~H.~Brandenberger,Phys.\ Rev.\ D {\bf 63}, 123501
(2001),   hep-th/0005209; 
R.~H.~Brandenberger and J.~Martin,Mod.\ Phys.\ Lett.\ A {\bf 16}, 999
(2001),  astro-ph/0005432.

\bibitem{N}
J.~C.~Niemeyer, Phys.\ Rev.\ D {\bf 63}, 123502 (2001), astro-ph/0005533;
J.~C.~Niemeyer and R.~Parentani, Phys.\ Rev.\ D {\bf 64}, 101301
(2001),  astro-ph/0101451.

\bibitem{stringscale} V.S. Kaplunovsky, Nucl. Phys. {\bf B307} (1988) 145,
hep-th/9205070; Erratum: ibid. {\bf B382} (1992) 436.

\bibitem{adds} N. Arkani-Hamed, S. Dimopoulos, G. Dvali,
Phys. Lett. {\bf B429} (1998) 263;
I. Antoniadis, N. Arkani-Hamed, S. Dimopoulos, G. Dvali,
Phys. Lett. {\bf B436} (1998) 257.
G. Shiu and S.-H.H. Tye, Phys. Rev. {\bf D58} (1998)
106007.


\bibitem{Kempf}  A. Kempf, Phys.\ Rev.\ D {\bf 63}, 083514 (2001),
 astro-ph/0009209.

\bibitem{KN}
A.~Kempf and J.~C.~Niemeyer, Phys.\ Rev.\ D {\bf 64}, 103501 (2001),
astro-ph/0103225.



\bibitem{EGKS}
R.~Easther, B.~R.~Greene, W.~H.~Kinney and G.~Shiu, 
Phys.\ Rev.\ D {\bf 64}, 103502 (2001)
hep-th/0104102.


\bibitem{hui01} L.~Hui and W.~H.~Kinney, astro-ph/0109107.


\bibitem{Shiu:2001sy}
G.~Shiu and S.~H.~Tye,
Phys.\ Lett.\ B {\bf 516}, 421 (2001).


\bibitem{MFB}
V.~F.~Mukhanov, H.~A.~Feldman and R.~H.~Brandenberger,
Phys.\ Rept.\  {\bf 215}, 203 (1992).


\bibitem{Easther} J.~Adams, B.~Cresswell and R.~Easther, \\
astro-ph/0102236.

\bibitem{Lambert} R.~Corless, G.~Gonnet, D.~Hare, D.~Jeffrey, and
D.~Knuth, {\it Adv. Comp. Math.\/} {\bf 5} (1996) 329.


\bibitem{sur}G. Venezario, Europhys. Lett. {\bf 2} (1986) 199;
        D. Gross and P. Mende, Nucl. Phys. {\bf B303} (1988) 407;
        D. Amati, M. Ciafaloni and G. Veneziano,
        Phys. Lett. {\bf B216} (1989) 41;
R. Guida, K. Konishi and P. Provero, Mod. Phys. Lett. {\bf A6} (1991)
1487.

\bibitem{abramowitz} M. Abramowitz and I.A. Stegun,
{\it ``Handbook of Mathematical Functions with Formulas, Graphs and
Mathematical Tables''}, Dover, New York, (1965).


\bibitem{turner} M.~S.~Turner, M.~White, and J.~E.~Lidsey,
Phys. Rev. D. {\bf 48}, 4613 (1993).

\bibitem{CMNP}
A.~G.~Cohen, G.~W.~Moore, P.~Nelson and J.~Polchinski,
Nucl.\ Phys.\ B {\bf 267}, 143 (1986).


\end{thebibliography}
\end{document}